\documentclass[12pt]{article}
\usepackage{graphicx}
\newcommand{\be}[3]{\begin{equation}  \label{#1#2#3}}     % non-hyper
\newcommand{\bib}[3]{\bibitem{#1#2#3}}
\newcommand{\ee}{ \end{equation}}
\newcommand{\ba}{\begin{array}}
\newcommand{\ea}{\end{array}}
\newcommand{\p}{\partial}
\newcommand{\NP}[3]{{\em Nucl. Phys.}{ \bf B#1#2#3}}
\newcommand{\PRD}[2]{{\em Phys. Rev.}{ \bf D#1#2}}
\newcommand{\PRL}[2]{{\em Phys. Rev. Lett.}{ \bf #1#2}}

\newcommand{\PL}[3]{{\em Phys. Lett.}{ \bf B#1#2#3}}                        
\renewcommand{\arraystretch}{1.7}
\font\mybb=msbm10 at 12pt
\def\bb#1{\hbox{\mybb#1}}

\def\R {\bb{R}}
\def\E {\bb{E}}
\def\M {\bb{M}}

 \def\unit{\hbox to 3.3pt{\hskip1.3pt \vrule height 7pt width .4pt \hskip.7pt
\vrule height 7.85pt width .4pt \kern-2.4pt
\hrulefill \kern-3pt
\raise 4pt\hbox{\char'40}}}

\begin{document}

\pagestyle{empty}
\rightline{UG-2/98}
\rightline{HUB-EP-14/98}
\rightline{hep-th/9803090}
\rightline{March 1998}
\vspace{2truecm}
\centerline{\Large \bf  $D$-Instantons and asymptotic geometries}
\vspace{2truecm}
\centerline{\bf E.~Bergshoeff}
\vspace{.5truecm}
\centerline{Institute for Theoretical Physics}
\centerline{Nijenborgh 4, 9747 AG Groningen}
\centerline{The Netherlands}
\vspace{1truecm}
\centerline{\bf K.\ Behrndt }
\vspace{.5truecm}
\centerline{Institut f\"ur Physik }
\centerline{Invalidenstra\ss{}e 110, 10115 Berlin}
\centerline{Germany}
\vspace{2truecm}
\centerline{ABSTRACT}
\vspace{.5truecm}

The large N limit of D3-branes is expected to correspond to a
superconformal field theory living on the boundary of the anti-de
Sitter space appearing in the near-horizon geometry.  Dualizing the
D3-brane to a D-instanton, we show that this limit is equivalent to a
type IIB S-duality. In both cases one effectively reaches the
near-horizon geometry.  This provides an alternative approach to an
earlier derivation of the same result that makes use of the properties
of a gravitational wave instead of the D-instanton.

\vfill\eject
\pagestyle{plain}
 
\noindent{\large \bf 1. Introduction}
 
\vspace{.5cm}

Ten years ago a relation was suggested between branes, singletons and 
anti-de Sitter ($adS$) space time in the context of the
``Brane living at the End of the Universe'' programme \cite{500,600,601,602}.
In this approach a brane is considered in an $adS$ background\footnote{
Such backgrounds are natural to consider since they occur in the
spontaneous compactification of supergravity theories \cite{500,600}.}
with the worldvolume of the brane positioned at the boundary of the
$adS$ space time. The suggestion was made that the degrees of freedom
of this brane are the singleton representations of the $adS$ group
and products thereof. These so-called singletons do not correspond
to local degrees of freedom in the bulk but instead describe boundary
degrees of freedom \cite{603}. At the same time the anti-de Sitter 
isometry group of the bulk manifests itself as a conformal group on the 
boundary of the $adS$ space time.
It therefore seems that the physics of the brane is 
determined by a conformal field theory defined on the $adS$ boundary.

Recently, there has been a renewed interest in these interconnections
from different points of view
% \cite{620,621,622,812,820,623,629,624,625,626,627}
\cite{620} -- \cite{628}.  One of the
observations is that the anti-de Sitter space time also occurs in
the (non-singular) near-horizon geometry of the ten-dimensional
D-3-brane and the eleven-dimensional M-2-brane and M-5-brane
\cite{845}. Recently, it has been suggested that the large N limit of 
D3-branes corresponds to a superconformal field theory living on the
boundary of this anti-de Sitter space \cite{621}.
Moreover, it was observed \cite{812,820} that via a
series of duality transformations the M-2-, D-3- and M-5-branes (with
flat asymptotic geometry) can be locally\footnote{ The global validity
of these duality transformations should be taken with caution
\cite{820}, see also the conclusions.}  transformed into a non-flat
geometry of the type $AdS_4 \times S_7, AdS_5 \times S_5$ and $AdS_7
\times S_4$, respectively. These geometries are exactly the
(non-singular) near-horizon geometries of the original brane. Notice
that the duality transformations have changed the asymptotic geometry.
These results are another hint that the physics of these branes are
described by supersingleton field theories \cite{820}. It is the
purpose of this letter to give an alternative derivation of the
duality symmetries relating branes to $adS$ spaces.

The basic idea of \cite{812,820} is to start from a brane solution 
(with $p$ spacelike isometries) described by a harmonic function
on the transversal space:
\be001
H = h + {Q\over r^{7-p}}\, ,\hskip 1truecm r^2 = x_{p+1}^2 + \cdots x_9^2\, .
\ee
Here $h$ is an integration constant and $Q$ represents the charge of
the brane. In order to have asymptotic flat geometries we will
restrict ourselves to p-branes with $p<7$. 
One first relates the brane solution,
via U-duality, to a gravitational wave solution
\be002
ds^2_{10} = dudv + Hdu^2 + dx_i^2\,, \hskip 1truecm i=2,\cdots 9,
\ee
where $(u,v)$ are lightlike coordinates parametrizing a
two-dimensional subspace, with signature (1,1), of the
ten-dimensional space time and $H$ is a harmonic function of
the eight-dimensional transversal space.
One next makes a change of coordinates that amounts to an $SL(2,\R)$
rotation in the $(u,v)$ space given by
\be160
\pmatrix{v \cr u } \rightarrow \pmatrix{ 1 & - h \cr 0 & 1 } 
\pmatrix{v \cr u }\, .
\ee
After this coordinate transformation one ends up with the same
gravitational wave solution but with the constant $h$ in the harmonic
function $H$ set equal to zero. Finally, one dualizes the wave back to
the brane solution one started from. The net effect of this web of
dualities is that one obtains the same brane solution but with the
constant in the harmonic set equal to zero.  This new solution
describes exactly the same geometry that one obtains upon approaching
the horizon of the original brane solution at $r=0$ since in that
limit one can effectively ignore the constant part in the harmonic
function.

Another way of shifting away the constant part of the harmonic
function has been discussed in \cite{830}. The basic idea here is to
relate the brane to a Kaluza-Klein (KK) monopole instead of a
gravitational wave.  One next considers the four-dimensional Taub-NUT
space of the KK monopole.  A $TST$ duality transformation\footnote{
This duality transformation has the same effect as an Ehlers
transformation \cite{830}.}  removes the constant part in the harmonic
and one ends up with an Eguchi-Hanson instanton.  Dualizing back to
the original brane leads to the same result as above.

In this letter we want to consider another intermediate solution which
has the advantage that it has a manifest $SL(2,\R)$ duality symmetry and that
the process of shifting away the constant part of the harmonic
functions has a simple interpretation as performing a special
$SL(2,\R)$ transformation. Since Type IIB superstring theory has a
manifest $SL(2,\R)$ symmetry it is natural to consider a IIB brane. In
fact, the most natural one to consider is the one with the highest
dimensional transversal space which is the D-instanton \cite{700}. We
therefore propose to use the D-instanton as an intermediate solution.

As explained in \cite{810} the $D$-instanton can be understood as
a compactified twelve-dimen\-sional wave. The metric of such a gravitational
wave is given by\footnote{The metric is in (twelve-dimensional) Einstein 
frame. Note that there is no dilaton in twelve dimensions.}
\be150
ds^2_{12} = dudv + H du^2 + ds_E^2\, ,
\ee
where $(u,v)$ are lightlike coordinates parametrizing a
two-dimensional torus with (1,1) signature and fixed volume. The
function $H$ is a harmonic function of the Euclidean ten-dimensional
space with metric $ds_E^2$. As shown in \cite{810} reducing the wave
(\ref{150}) over $(u,v)$ yields the $D$-instanton solution. From this
twelve-dimensional point of view the special $SL(2,\R)$ transformation
that transforms away the constant part of the harmonic function
describing the D-instanton corresponds to an $SL(2, \R)$ rotation in
the $(u,v)$ space as given by (\ref{160}).  The difference with the
approach of \cite{812,820} is that in that case the $SL(2,\R)$
rotation is performed on a two--dimensional subspace of the
ten-dimensional space time.  Here, however, we rotate the two
additional dimensions arising in a twelve-dimensional interpretation
of type IIB superstring theory \cite{825,826}. This rotation can be
interpreted as a special SL(2,\R) duality transformation of the Type IIB
superstring theory.

In the next Section we first review the D-instanton solution.
In Section 3 we describe the web of dualities that makes use of the
D-instanton. The supersymmetry of the different configurations, before
and after duality, is considered in Section 4 while the extension to
intersecting configurations is discussed in Section 5. A further
discussion and interpretation of our results can be found in the
Conclusions.

\bigskip

\noindent{\large \bf 2. The $D$-Instanton}

\medskip

\noindent

In this section we review the properties of the $D$-instanton solution
\cite{700} of IIB supergravity \cite{701}.  Since all gauge fields
vanish for this solution we only consider the Ramond/Ramond (RR)
pseudo--scalar $\ell$, the dilaton $\phi$ and the metric $g_{\mu\nu}$.
Introducing the complex scalar $S$ via
\be010
S = \ell + i e^{-\phi}\, ,
\end{equation}
the Minkowskean IIB action in Einstein frame  is given by
\be020
\ba{rcl}
S &=& \int d^{10}x \sqrt{|g|} \left[R + {1\over 2}{\partial S \partial 
\bar S \over 
({\rm Im}S)^2} \right] + S_{\p M} \\
&=& \int d^{10}x \sqrt{|g|} \left[ R + {1 \over 2} 
\biggl (e^{2 \phi} (\partial \ell)^2 \, 
 + (\partial \phi)^2\biggr )
\right] + S_{\p M} \ ,
\ea
\end{equation}
where $S_{\partial M}$ is a boundary contribution that
will be discussed in the Conclusions.
After performing a Wick rotation to Euclidean space the action reads
\be030
\ba{rcl} \label{E}
S_E &=& \int d^{10}x \sqrt{|g|}\left[ - R + {\partial S_+ \partial S_- 
\over {1 \over 2} (S_+ - S_-)^2} \right] + S_{\partial M} \\
&=& \int d^{10}x \sqrt{|g|}\left[- R + {1 \over 2}\biggl ( 
e^{2 \phi} (\partial \ell)^2 - (\partial \phi)^2\biggr ) 
\right] + S_{\partial M} \ ,
\ea
\end{equation}
where the two real scalars $S_{\pm}$ are defined as
\be040
S_{\pm} = \ell \pm e^{-\phi}\, .
\end{equation}
Note, that the kinetic term of $\ell$ has changed 
its sign, due 
to the fact that $\ell$ is a pseudo--scalar, which under a Wick
rotation\footnote{Both the Wick rotation and the parity transformation
can be viewed as special cases of a continuous phase transformation,
as has been discussed in \cite{800}.  From this point of view, the Wick
rotation can be seen as the square root of a parity transformation and 
therefore $\ell$ transforms as $\ell \rightarrow i\ell$.}
transforms as $\ell \rightarrow i\ell$. 

The equations of motion corresponding to the Euclidean action (\ref{E})
are given by
\be050
\begin{array}{rcl}
R_{\mu\nu}&=& e^{2 \phi} \partial_{\mu} \ell \partial_{\nu} \ell - 
\partial_{\mu} \phi \partial_{\nu} \phi\, , \\
0&=&\partial_{\mu}\left(\sqrt{|g|}\, g^{\mu\nu} e^{2\phi} \,
\partial_{\nu} \ell \right)\, ,\\
0&=&e^{2\phi} (\partial \ell)^2 + {1 \over \sqrt{|g|}}\, 
\partial_{\mu} \left(\sqrt{|g|} \, g^{\mu\nu}\partial_{\nu}\phi \right)\, .
\end{array}
\end{equation}

We now search for solutions to these equations of motion. Following 
\cite{700} we assume that the Einstein metric describing the 
ten--dimensional space transverse to the D--instanton is flat, i.e.
\be055
ds^2 = dr^2 + r^2 d\Omega_9,
\ee
where we have used spherical coordinates\footnote{
The results of
\cite{801,627} suggest that there exists a more general class of
D--instanton solutions which can be obtained by replacing the
metric $d\Omega_9$ of the 9-spere in (\ref{055}) by the metric
$ds^2_{\rm comp}$ describing the geometry of any Einstein space that arises
in the compactification of (Euclidean) 
IIB supergravity from 10 to 1 dimensions.
The 1-dimensional space has an $\R^+$ topology and is parametrized by the
radial coordinate $r$. We will
not consider this possibility further in this work and restrict ourselves
to the standard D-instanton with flat transverse space.}.
Under this assumption the general solution to the equations of motion
(\ref{050}) is given by
\be060 
\pm \ell + \alpha = e^{-\phi} = {1 \over H}\, , 
\end{equation}
where $\alpha$ is constant and $H$ is a general harmonic function over
the 10-dimensional flat Euclidean space, i.e.\ $\partial^2 H
=0$. For the spherical symmetric case this harmonic
function is given by
\be070
H = h + {Q \over r^8} \ ,
\end{equation}
where $h$ is an integration constant and $Q$ is the Noether charge
defined by \cite{700}
\be080
Q = \pm {1 \over 8 \Omega_9} \int_{\partial M} e^{2\phi} \, \partial \ell\, ,
\end{equation}
where $\Omega_9 = {2\pi^{5/2} \over 24}$ is the volume of the 9-sphere.
Therefore, in this case the D-instanton solution is parameterized 
by the three constants $\alpha, h$ and $Q$. 

In the string frame the D-instanton solution (\ref{060}) reads
\be090
\begin{array}{l}
\pm \ell + \alpha = e^{-\phi} = H^{-1}\, , \\
ds^2 = \sqrt{H} \left[dr^2 + r^2 d\Omega_9 \right] = 
 \sqrt{h r^4 + {Q \over r^4}} \left[\left(dr \over r\right)^2 + 
d\Omega_9 \right]\, . 
\end{array}
\end{equation}
Note that the solution is symmetric under the interchange
\be100
h r^4 \ \leftrightarrow \ {Q \over r^4}\, . 
\end{equation}
It corresponds to a wormhole connecting two asymptotic flat regions
(see Figure 1).
The minimal diameter $d_{min}$ of the wormhole--throat equals 
\be101
d^8_{min} =  32^2hQ
\end{equation}
and it is positioned at a value $r=r_{min}$ given by
\be102
r_{min}^8 = {Q \over h} \ .
\ee
Under the ``mirror'' symmetry (\ref{100}) the asymptotic flat regions
at $r=0$ and $r=\infty$ are mapped onto each other, while at the same
time $Q$ and $h$ get interchanged.  Notice however that, although the
metric is symmetric under this ``mirror'' symmetry, the dilaton is
not. In the $r= \infty$ vacuum the dilaton is finite whereas it
diverges for $r \rightarrow 0$.  The asymptotic geometry at $r=0$ is
given by a flat space time with metric (in string frame)
\be104
ds^2 =  \sqrt{{Q \over r^4}} \left[\left(dr \over r\right)^2 + 
d\Omega_9 \right] = d\rho^2 + \rho^2 d\Omega_9\, ,
\ee
where $\rho = Q^{1 \over 4}/r$. 
\begin{figure} \label{fig1}
\begin{center}
%\includegraphics[angle=0, width=40mm]{instanton.eps} 
%\qquad
\includegraphics[angle=0, width=70mm]{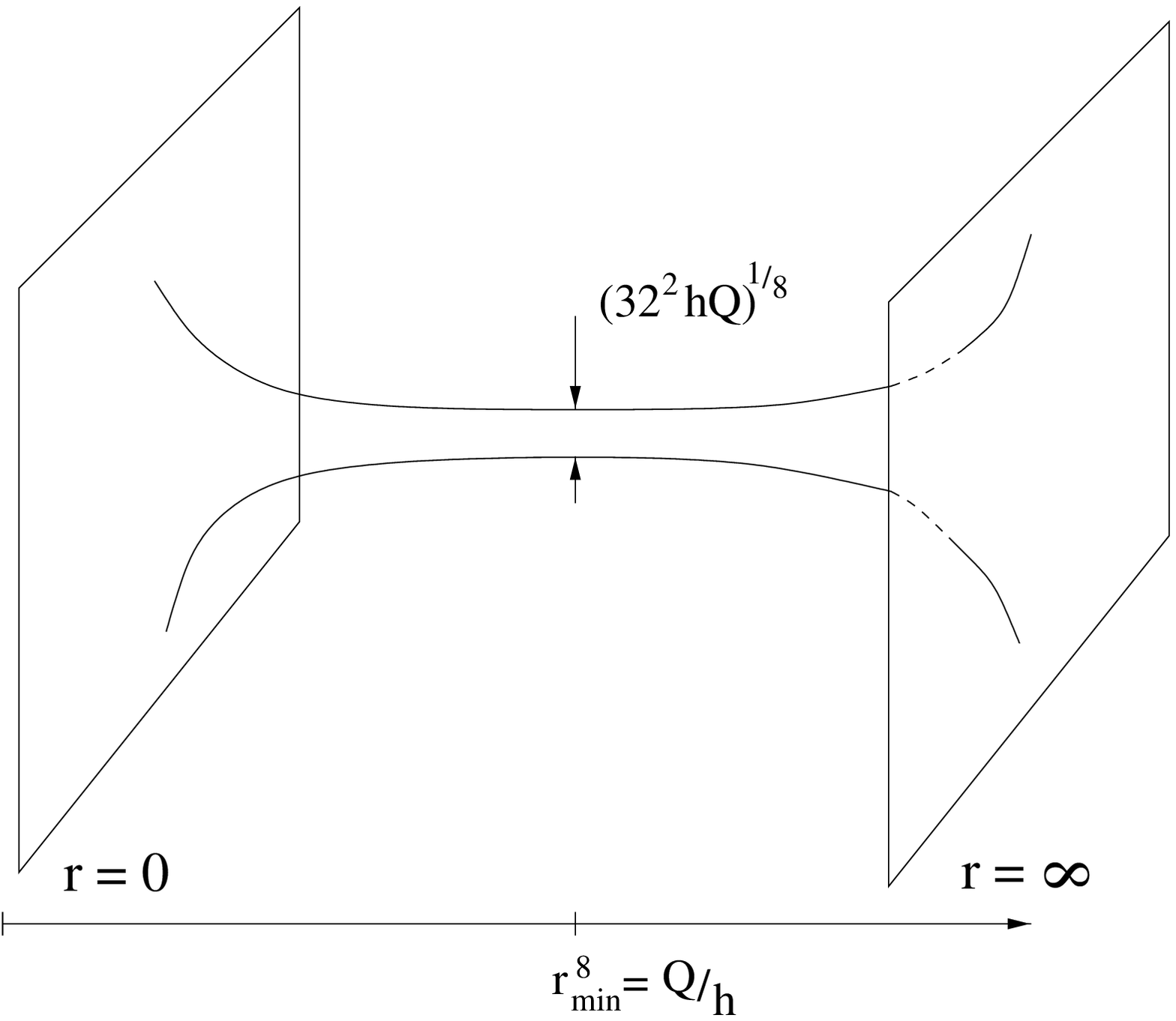}
\end{center}
Figure 1: {\small {\bf The D-instanton geometry}: The D-instanton solution
comprises two asymptotic flat regions,
one at $r=\infty$ and one at $r=0$ which are connected by a throat with minimal
diameter $d_{min} = (32^2hQ)^{1/8}$ at position $r_{min} = (Q/h)^{1/8}$.}

\end{figure}

The equations of motion (\ref{050}) are invariant under the
$SL(2,\R)$ transformations
\be110
S_{\pm} \rightarrow {a S_{\pm} + b \over c S_{\pm} +d } \qquad , 
\qquad ad-bc =1\, ,
\end{equation}
with the two generators
\be120
\Omega_1: \ S_{\pm} \rightarrow S_{\pm} + 1 \qquad  , \qquad 
\Omega_2: \ S_{\pm} \rightarrow - {1 \over S_{\pm}}\, .
\end{equation}
Using these transformations we can change arbitrarily the parameters
$\alpha$ and $h$ characterizing the D-instanton solution
but {\em not} $Q$. Especially we can transform the
constant part $h$ of the harmonic function (\ref{070}) to zero.
Taking the positive sign in (\ref{060}) this is achieved by the 
special $SL(2, \R)$ transformation\footnote{
A similar transformation exists for the negative sign in (\ref{060}).
}
\be130
\pmatrix{a & b \cr c & d} = \Omega_1^{-\alpha} \Omega_2^{-1} 
(\Omega_1)^{h/2} \Omega_2 \Omega_1^{\alpha} 
= \pmatrix{ 1 + {h\alpha \over 2}  & {\alpha^2 h \over 2} \cr 
-{h \over 2} & 1- {\alpha h \over 2}}\, ,
\end{equation}
yielding the solution
\be140
\ell + \alpha = e^{-\phi}  = {r^8 \over Q}\, .
\end{equation}
The special $SL(2,\R)$ transformation (\ref{130}) will play an
important role in the next Section.  Notice that the above solution
can also be obtained by performing a special dimensional reduction of
the twelve--dimensional wave which is different from the one discussed
in the Introduction \cite{Strings97}.

Given the expression (\ref{102}) for the position $r_{min}$ of the
wormhole we see that by changing the value of $h$ we effectively move
the position $r_{min}$.  In other words, due to the $SL(2, \R)$
symmetry, the position left or right from the wormhole is not
determined. Any point positioned at $r_{fix}$ with $r_{fix} > r_{min}$
is $SL(2, \R)$ equivalent to a point positioned at $r_{fix}$ with
$r_{fix} < r_{min}$ and especially by transforming $h$ away to $h=0$
one effectively moves towards the vacuum at $r=0$ where one reaches
the flat space time, see eq.\ (\ref{104}).

\bigskip

\noindent
{\large \bf 3. $T$-duality}

\medskip

In the previous section we have shown that the constant part $h$ of
the harmonic function $H$ can be removed by a special $SL(2, \R)$
transformation. This has important consequences for the geometry
described by the solution.  For $h\ne 0$, the solution represents a
wormhole geometry, see Figure 1, whereas for $h=0$ we obtain a flat
space time. This flat space time geometry can be parametrized
by different coordinate systems ($r$ versus $\rho$), see eq.\
(\ref{104}).  We therefore have three (locally) equivalent
representations of the $D$-instanton: 

(a) the original D-instanton solution (\ref{090}); \newline
(b) the flat space time geometry (\ref{104}) in
 the $r$ coordinate system; \newline
(c) the flat space time geometry (\ref{104}) in the 
 $\rho$ coordinate system.

In this section we will consider the $T$-dual versions of these three
different representations.

{\sl Representation (a)}\ \ \ 
 We start with the ``standard'' representation given in
(\ref{090}). Since there is no worldvolume direction we can only
apply a T-duality in the transversal space\footnote{
Actually, following \cite{628} one can also perform a 
so-called ``Hopf T-duality''. In order to perform this kind of duality,
one first parametrizes the 10-dimensional transversal space of the
D--instanton in polar
coordinates, see (\ref{055}). One next realizes the 9-sphere $S^9$ as
a $U(1)$ bundle over $CP^4$ and performs a T-duality in the $U(1)$ isometry
direction. One thus obtains a D-0-brane with a non-flat transverse
space involving the $CP^4$ manifold. 
}. It is well-known that, after applying T-duality in
the different transversal directions, 
one obtains the D-p-branes ($0\le p \le 6$) with
string-frame metric given by\footnote{ Note that after performing the
$T$-duality we rotate the Euclidean space back to a Minkowskean space
time.}
\be170
ds^2 = {1 \over \sqrt{H}}dx^2_{p+1} - \sqrt{H} dx_{9-p}^2 \quad , \quad 
e^{-2\phi} = H^{p-3 \over 2} \quad , \quad F_{0..pI} = \p_I H^{-1} \, , 
\ee
where $I= p+1,\cdots ,9$ represents the $9-p$ transverse directions.
The function $H$ is harmonic only with respect to the 
transverse directions, i.e.
\be171
H = h + {Q\over r^{7-p}}\quad , \hskip 1.5truecm 
r^2 = x_{p+1}^2 + \cdots +  x_9^2\, .
\ee
\bigskip

{\sl Representation (b)}\ \ \ 
 Next, we consider the $T$-dual version of the second
representation of the D-instanton, i.e.~the flat space time geometry
in the $r$ coordinate system, see (\ref{104}).  This obviously leads
to the same $D$-brane configuration given above but now with the
constant part $h$ in the harmonic function $H$ being removed.  On the
other hand, approaching the horizon of a D-p-brane solution at $r=0$
one can effectively ignore the constant part in the harmonic function.
Therefore $T$-dualizing (\ref{104}) (in the $r$-basis) yields the
near-horizon geometry of the D-p-brane and therefore by the $SL(2,\R)$
transformation (\ref{130}) we effectively have approached the horizon.  The
solution in string-frame is given by
\be180
\ba{l}
ds^2 = \sqrt{r^{7-p} \over Q} dx^2_{p+1} - \sqrt{Q \over r^{3-p}}
\left[ \left({dr \over r}\right)^2 + d\Omega_{8-p}\right]\, , \\
e^{{-4 \over p-3}\phi} = {Q \over r^{7-p}} \quad , \quad F_{0..pI} = 
\p_I ({r^{7-p} \over Q}) \ .
\ea
\ee
Note, that for all $p$ the spherical part of the above solutions is
singular, except for $p=3$, where one obtains the $AdS_5 \times S_5$
geometry. Thus, the D-3-brane interpolates between the Minkowkean
vacuum ($r=\infty$) and the $AdS_5 \times S_5$ vacuum ($r=0$)
\cite{845} in the same way as the D-instanton interpolates between the
two Minkowskean vacua at $r=0$ and $r=\infty$ and both regions
are interchanged by the mirror transformation (\ref{100}).
\bigskip

{\sl Representation (c)}\ \ \ 
Finally, we consider the $T$-dual version of the third
representation of the D-instanton, i.e.~the same flat space time
geometry as above but now in the $\rho$ basis, see (\ref{104}). More
precisely, starting from the $\rho$ basis, we first introduce
Cartesian coordinates and next apply $T$-duality.  Since we are
dealing with a flat metric, the only changes are in the gauge
fields. We thus arrive at the following string-frame configuration
($0\le p\le 6$)
\be190
ds^2 = dx^2_{p+1} - dx^2_{9-p}\  ,  \qquad
e^{{-4 \over p-3}\phi} = Q^{p-3 \over 4} \rho^{7-p}\ ,  \quad F_{0..pI} = 
 Q^{p-3 \over 4} \p_I \rho^{7-p}\, ,
\ee
where $\rho$ is the radial coordinate of the transversal part.  These
configurations differ from the one given in (\ref{180}).  This was to
be expected since $T$-duality and coordinate transformations do not
commute. It is straightforward to verify that the above configuration
indeed solves the equations of motion: the (p+1)-form gauge fields
solve the free Maxwell-like equations of motion and the expression for
the dilaton is just the solution of a Laplacian equation.  Notice that
the relative factors between the scalar and gauge field part of the
solution ensure that the energy momentum tensor vanishes as it has to
be for a flat metric.

\pagebreak

\bigskip

\noindent
{\large \bf 4. Supersymmetry}
\medskip

In this section we consider the supersymmetry of the D-instanton
solution and its $T$-dual versions. Ignoring the gauge fields, the IIB
supersymmetry rules of the gravitino and dilatino (using the Einstein
metric) in Minkowskean space time are given by\footnote{These
supersymmetry rules, using an $SU(1,1)$-basis, have been given in
\cite{701}. We use here the $SL(2,\R)$-covariant form of these rules,
as given in \cite{850}.}
\be200
\ba{l}
\delta \psi_{\mu} = \left(\p_{\mu} - {1 \over 4} \omega^{ab}_{\mu} 
\Gamma_{ab} - {i \over 4} \, e^{\phi} \, \p_{\mu} \ell \right) 
\epsilon\, ,  \\
\delta \lambda = {1 \over 4} \Gamma^{\mu} \epsilon^{\star} 
\left(\p_{\mu} \phi + i \, e^{\phi} \p_{\mu} \ell \right)\, .
\ea
\ee
The supersymmetry of the D-instanton solution (\ref{060}) has already been
considered in \cite{700}.
After a Wick rotation to a ten-dimensional Euclidean space
the supersymmetry transformations become (in Einstein frame) \cite{700}
\be210
\ba{l}
\delta \psi_{\mu}^{(\pm)} = \left(\p_{\mu} - {1 \over 4} \omega^{ab}_{\mu} 
\Gamma_{ab} \mp {1 \over 4} \, e^{\phi} \, \p_{\mu} \ell \right) 
\epsilon^{(\pm)}\, ,  \\
\delta \lambda^{(\pm)} = {1 \over 4} \Gamma^{\mu} \epsilon^{(\mp)} 
\left(\p_{\mu} \phi \pm e^{\phi} \, \p_{\mu} \ell \right)\, .
\ea
\ee
Inserting the solution (\ref{060}) with the ``+'' sign and taking into 
account that the Einstein metric is flat, one obtains as a solution
\be220
\epsilon^{(+)} = 0\  , \qquad \qquad 
\epsilon^{(-)} = e^{\phi \over 4} \, \epsilon_0^{(-)}  = 
H^{1\over 4} \, \epsilon_0^{(-)} 
\ee
for constant spinor $\epsilon_0^{(-)}$. For the negative sign
in (\ref{060}) one obtains a similar solution, where 
$\epsilon^{(+)}$ and $\epsilon^{(-)}$ are interchanged.
Therefore, the D-instanton
solution generically breaks 1/2 of the supersymmetry. The same is true
for the $SL(2, \R)$ transformed solution (\ref{140}) since the
supersymmetry rules (\ref{210}) are $SL(2, \R)$-covariant.

Next, we discuss the supersymmetry in the two asymptotic flat
regions. First, in the $r= \infty$ vacuum obviously all supersymmetry
is restored ($\phi = \ell =$ constant). 
 Requiring that the gravitino
  variation vanishes we find that the spinors in the limit $r
 \rightarrow 0$ behave like $\epsilon^{(\pm)} \sim r^{2}
 \epsilon_0^{(\pm)}$. 
Thus, both spinors $\epsilon^{(\pm)}$ vanish like $r^2$
and as a consequence both dilatino variations $\delta \lambda^{(\pm)}$
vanish identically.  We conclude that, 
{\em in both asymptotic
regions of the D-instanton solution ($ r= \infty$ and $r=0$) we have a
restoration of unbroken supersymmetry}.

Next, we consider the supersymmetry of the $T$-dual versions of the
D-instanton.  After $T$-duality and rotating back to Minkowskean
signature the relevant part of the supersymmetry variations in the
string frame become ($0\le p \le 6$) \cite{851}\footnote{Actually,
for p=3, due to the selfduality condition of the 5-form field-strength,
one should include an extra factor of 1/2 in front of the
$\Gamma\cdot F$ term in the gravitino rule.
We thank Kostas Sfetsos for pointing this out to us.} 
\be230
\ba{l}
\delta \psi_{\mu} = \p_{\mu} \epsilon - {1 \over 4} \omega^{ab}_{\mu}
\Gamma_{ab}\epsilon + {(-)^p \over 8 (p+2)!} \, e^{\phi} \, 
(F\cdot \Gamma) \, \Gamma_{\mu}\epsilon_{(p)}'\, ,  \\
\delta \lambda = {1\over 4} \Gamma^{\mu}   (\p_{\mu} \phi) \epsilon^{\star}+ 
{3-p \over 16 (p+2)!} \, e^{\phi} \,  (F \cdot \Gamma)  
\epsilon_{(p)}^{\prime \; \star}\, ,
\ea
\ee
where $(F \cdot \Gamma) = F_{\mu_1 \cdots \mu_{p+2}} \Gamma^{\mu_1
\cdots \mu_{p+2}}$ and the parameters $\epsilon'_{(p)}$ are defined in
Table 1.
\renewcommand{\arraystretch}{1.3}

\begin{center}
\begin{tabular}{|c|c|c|c|}
\hline
$p$&$\epsilon_{(p)}^\prime$\ (IIA) & $p$&$\epsilon_{(p)}^\prime$\ (IIB)\\
\hline\hline
$0$ &$\epsilon$&$1$&$i\epsilon^\star$\\
$2$&$\gamma_{11}\epsilon$&$3$&$i\epsilon$\\
$4$&$\epsilon$&$5$&$i\epsilon^\star$\\
$6$&$\gamma_{11}\epsilon$&$-$&$-$\\
 \hline
 \end{tabular}
\bigskip
\end{center}

\nopagebreak
Table 1: {\small {\bf Definition of spinors $\epsilon_{(p)}^\prime$}:
     The table gives the definition of the spinor parameters 
     $\epsilon_{(p)}^\prime$ occurring in the supersymmetry transformations   
     (\ref{230}) in terms of the supersymmetry parameter $\epsilon$.}

\bigskip

Substituting the D-p-brane solutions (\ref{170}) or the solutions
(\ref{180}) into the supersymmetry rules (\ref{230}) 
we find as solution for the vanishing of these supersymmetry
variations
\be240
\epsilon + \Gamma_{0 \cdots p} \epsilon_{(p)}' = 0\, ,
\qquad  \qquad
\epsilon = H^{- {1 \over 8}} \epsilon_0\, .
\ee
This shows that the solutions (\ref{170}) and (\ref{180}) for all $p$
have half of unbroken supersymmetry.  In addition for special cases we
have a restoration of unbroken supersymmetry \cite{847}.  For the
D-p-brane solutions (\ref{170}) this is the case in the asymptotic
vacuum ($r = \infty$) and for the 3-brane case this happens also near
the horizon ($r=0$). For the solutions (\ref{180}) we only have
unbroken supersymmetry for $p=3$ in which case the solution has the
$AdS_5 \times S_5$ geometry.

Finally, we consider the supersymmetry of the solutions
(\ref{190}). Although it seems to be natural to take this coordinate
system, it has important consequences for the supersymmetry.  Considering
the ``worldvolume'' components (= isometry directions) of the
gravitino variation (\ref{230}), we find that for $0\le p \le 6$ all
supersymmetries are broken.  The technical reason for this is that in the
worldvolume components of the
gravitino variation the $\partial\epsilon$ vanishes since the Killing 
spinor does not depend on the worldvolume
directions and  furthermore without gravity we have that $\omega_{\mu}^{ab}$
vanishes as well. However,
the field strength $F$ is non-trivial and this leads to a complete
breaking of the supersymmetry.  This result might be
surprising since the original ($p=-1$) solution allowed 1/2 of
unbroken supersymmetry. However this case is special since, although
there is no gravity, neither are there any ``worldvolume'' components
of the gravitino variation to consider.

\bigskip

\noindent
{\large \bf 5. Intersections}
\medskip

Turning on and off the constant parts in the harmonic functions has
also important consequences for intersections. Let us start with
presenting a systematic way of constructing intersections starting
from the $D$-instanton.  The idea is to solve the scalar field
equations for $\phi$ and $\ell$, that determine the instanton, in the
background of a further brane. We are interested in threshold bound
states (intersections) and therefore the gauge fields are given by
independent harmonic functions (or charges).  For the instanton this
means, that $\ell$ has to be independent of the background brane and
given the equation of motion for $\ell$ one realizes that only the
harmonic function of a 3-brane drops out\footnote{A D-7-brane
background is also possible, but in that case the harmonic
functions of the two intersecting branes
depend on different (relative transverse) coordinates. We
will not consider this possibility further here.}.

In the Einstein equations the instanton does not contribute to
the energy momentum tensor (due to the ansatz (\ref{060})) and thus,
they are solved by the D-3-brane metric, see eq.~(\ref{170}).
Next, using the instanton ansatz (with harmonic function $H_1$)
the two scalar field equations in (\ref{060}) become
\be250
0=\partial_{\mu}\left(\sqrt{|g|} \, g^{\mu\nu} \partial_{\nu} e^{\phi}
\right)\, ,
\ee
where we have to insert the D-3-brane metric (with harmonic function
$H_2$) in the Einstein frame.  For this brane however, the Einstein
and string metric are the same and we find $\sqrt{|g|}\, g^{\mu\nu} =
\delta^{\mu\nu}$. Hence we obtain a harmonic equation
\be260
e^{-\phi} = \pm \ell + \alpha_1 = {1 \over H_1} \ .
\ee
Notice that {\em only} for the D-3-brane (and D-7-brane) background we
get a flat Laplacian, for all other branes the scalar fields are more
complicated, indicating a violation of the harmonic superposition rule
\cite{855}. By this procedure, one naturally obtains the two known
families of intersections of two branes, which have 4 (or 8)
relative-transversal coordinates. Only in the first class
(corresponding to the D-3-brane case) both harmonic functions depend
on the overall transversal coordinates.

We thus obtain the intersection\footnote{After the submission of this
work we learned that a configuration describing a localized
D--instanton solution within a D-3-brane system has been obtained in
\cite{Tseytlin2}.} $(-1)\times 3$, from which we can obtain all other
intersecting brane configurations by $U$-duality. Approaching the
horizon the 4-d relative transversal space factorizes, i.e.\ $\M_{10}
\rightarrow \M_6 \times \E_4$.  We therefore end up effectively with a
6-d theory, with branes coming from 10-d intersections. Like in 10
dimensions also in 6 dimensions we have just one non-singular object,
the self-dual string.  Only for this object the space time factorizes
further into $AdS_3 \times S_3 \times \E_4$.  The 10-d configurations
giving this geometry are $1\times 5$, $2 \times 4$ and $3 \times 3$.

Like for the single branes one can reach the near-horizon geometry
also by an $SL(2, \R)$ transformation.  Applying the special
$SL(2,\R)$ transformation (\ref{130}) to the fields given in
(\ref{260}) one turns off the constant part of $H_1$.  This $SL(2,\R)$
transformation does not effect the other fields corresponding to the
D-3-brane since the D-3-brane is $SL(2,\R)$ invariant. In a second
step, using $U$-duality, we turn off the constant part of the second
harmonic function $H_2$ . By this procedure we obtain a 10-d space
time factorizing into $\M_{10} = \M_{6} \times \E_4$, where $\E_4$
denotes the relative transversal part. If the branes intersect over a
0- or 2-brane the $\M_6$ part does not factorize further, but if it
intersects over a string the space time factorizes further into
$\M_{10} = AdS_3 \times S_3 \times \E_4$.

As an example, consider the case $3 \times 3$ which is
given by
\be270
ds^2_{3\times 3} = {1 \over \sqrt{H_1 H_2}} \left( dt^2 - dz^2 \right) -
   \sqrt{H_1 H_2}(dx_m)^2 - \sqrt{H_1 \over H_2} (dx_6^2 + dx_7^2) -
\sqrt{H_2 \over H_1} (dx_8^2 + dx_9^2)\, ,
\ee
where $z$ denotes the direction of the common string,
$x_6, \cdots , x_9$ are the relative
transversal coordinates and the harmonic functions
are given by $H_i = 1 + q_i/r^2\ (i=1,2)$.
Using the $U$-duality chain described in Section 3
this configuration is locally for any radius $r$ equivalent to
\be280
ds^2_{3\times 3} = {r^2 \over \sqrt{q_1 q_2}} \left( dt^2 - dz^2 \right) -
   \sqrt{q_1 q_2}\left({dr \over r}\right)^2 - \sqrt{q_1 q_2} \, d\Omega_3  
- \sqrt{q_1 \over q_2} (\E_2) - \sqrt{q_2 \over q_1} (\E_2)
\ee
which is $AdS_3 \times S_3 \times \E_4$.

It is not difficult to add further branes to the intersecting
configuration.  We will start with 4 intersecting branes and
afterwards, by truncation, we will consider the triple intersections
as a special subclass.  The simplest way is to start with the $3
\times 3$ configuration (\ref{270}), to add a wave $W$ along the
common string direction $z$ and to insert a KK-monopole $KK$ with its
Taub-NUT part in the four-dimensional overall transversal space. This
intersection is $U$-dual to $(-1) \times 3 \times W \times KK$ and
because the 3-brane as well as the wave and KK monopole are $SL(2,
\R)$ invariant, we can again use the $SL(2, \R)$ rotation (\ref{130})
to turn off the constant part of the harmonic function describing the
$(-1)$-brane and subsequently we can do the same for all other branes
as well.  For the triple intersections we have to turn off either the
wave, obtaining the 5-d string, or the KK monopole yielding the 5-d
black hole.

In complete analogy to the double intersections discussed before, the
geometry factorizes. Keeping at least 3 overall transverse dimensions
the triple intersections factorize into $\M_{10} = AdS_3 \times S_2
\times \E_5$ for the 5-d string, $\M_{10} = AdS_2 \times S_3
\times \E_5$ for the 5-d black hole and finally the quadruple intersection
corresponding to the 4-d black hole factorizes into $\M_{10} = AdS_2
\times S_2 \times \E_6$. This leads exactly to the non-singular
near-horizon geometries considered in \cite{847,820}.

%%%%%%%%%%%%%%%%%%%%%%%%%%%%%%%%%%%%%%%%%%%%%%%%%%%%%%%%%%%%%%%%%

\bigskip

\noindent
{\large \bf 6. Conclusions}

\medskip

In this paper we have employed, via the D--instanton, the $SL(2,\R)$
duality symmetry of the Type IIB superstring theory to turn off the
constant part in the harmonic function describing brane solutions.
This is in analogy to the large N limit of the $D$-3-brane \cite{621},
in both cases the constant part is effectively neglected. This process
has important consequences for the geometry of the solution.  The
$D$--instanton geometry is a wormhole that interpolates between two
asymptotic flat space vacua and there is a mirror map (\ref{100}) that
transforms both vacua into each other.  As one can see in Figure 1 if
the constant part $h$ of the harmonic function vanishes, the throat
shrinks and the minimum moves towards infinity, i.e.\ the vacua at
infinity disappear and the throat closes. One ends up with a flat
space time, both in Einstein as well as in string frame metric.  Of
course, the same happens also in the mirror base $\rho$ defined below
(\ref{104}).  Notice, this situation (i.e.\ vanishing $h$) is reached
simply by employing a symmetry of the theory, namely the type IIB
$SL(2,\R)$ duality transformation (\ref{130}). We discussed three
topological different representations of the D-instanton (one wormhole
and two flat space descriptions), which are $SL(2,\R)$--dual to each
other.

By applying standard $T$-duality one can convert the $D$-instanton
into all other $D$-branes. Starting with the case of non-vanishing $h$
one obtains the standard branes, but if $h=0$ we had to distinguish
between the two mirror bases. In one case one gets the near-horizon
geometry of the $D$-branes, but in the other case one obtains flat space
time with non-trivial gauge fields and dilaton. Investigating the
supersymmetry we found, that in the first case 1/2 of supersymmetries
are broken and for the 3-brane all supersymmetries are restored, but
for the latter case (flat space time) all supersymmetries are
broken. The supersymmetry breaking in the latter case can be
understood due to the coordinate transformation (mirror map).

It is straightforward to apply the same procedure discussed above to
intersecting
branes and we discussed a procedure to construct intersecting
branes ``out of the $D$-instanton''. Thus, we could turn off all constant
parts by $S$-dualizing the $D$-instanton part. For all brane configurations
with a non-singular horizon this has the consequence that the space
time factorizes into the structure $AdS_p \times S_q \times \E_r$.
At this point we should stress, that in order to keep a well-defined
low-energy limit we should keep all charges sufficiently large
and a certain hierarchy to keep all higher derivative corrections
under control.

Summarizing, we are facing an intriguing situation.
In this work we have shown that by applying
$T$- and $S$-duality one can move effectively towards the
horizon, where for the non-singular cases the supersymmetry is
enhanced.  In addition, in case of the $D$--instanton we changed the
topology, simply by duality. But, as stressed at many places, all
discussions are valid only locally. We do not yet have a complete
picture about what happens globally, but there are some interesting
points to mention. A first hint comes from the action. The $\ell$
field is the gauge field for the $D$-instanton and we see, that if one
turns off the constant part of the harmonic function
the gauge coupling (which is the $e^{2\phi}$ factor in front of the $\ell$
kinetic term, see (\ref{030}))  vanishes at
infinity, indicating a phase transition. Geometrically, in this phase
transition one asymptotic flat region of the $D$-instanton disappears
or for the non-singular $D$-brane configurations the space time
factorizes.  This change on the boundary is also visible in
the $D$-instanton action, which appears by integrating out
the boundary terms in (\ref{030}) (see \cite{700}), and is given by
\be300
% S_{\p M} = \int_M d \left( e^{2\phi} \ell \; ^{\star}\wedge d \ell\right)
% =  {Q \over h} \, \Omega_9 \ ,
S_{\p M} = Q \, (\ell \pm e^{-\phi})_{\infty} \, \Omega_9 \ .
\ee 
Because $(e^{-\phi})_{\infty} = 1/h$ and if $h \rightarrow 0$, 
this term becomes infinite. This  was expected,
because instantons describe tunnel processes between the two
asymptotic vacua and they are exponentially suppressed in the
weak coupling limit, which is given by $h = (e^{\phi})_{\infty} = g
\rightarrow 0$. 

The procedure discussed in this paper supports the idea
of the holographic principle \cite{866}. Taking any non-singular brane
configuration, we were arguing, that any (finite) point in space time
is $U$-dual to the geometry $AdS_q \times S_p \times \E_r$.  This
space can be reduced to the anti-de Sitter space, which is  fixed
by a (conformal) field theory living on boundary \cite{621,623}.  This
means that the physics at any (finite) point is determined by the boundary and
not by local degrees of freedom. 

Finally, in this work we used  the ``standard''
D-instanton solution which has a flat transversal space. 
As suggested in \cite{801,627} there exists a more
general class of brane solutions, where the spherical part ($S_9$ in
our case) is replaced by a general Einstein space, see the footnote below
(\ref{055}).  By doing this one breaks in general
more supersymmetry and the spherical parameterization we used in this work
 is maximal
supersymmetric. The orbifold construction discussed in \cite{626}
is one example of this construction.
It would be of interest to extend the results of this work to these
more general D--instanton solutions.

\vspace{1truecm}

\noindent {\bf Acknowledgements}
\vspace{.2truecm}

E.B. thanks S.-J.~Rey for stimulating discussions and the physics department
of Seoul National University for its hospitality.
K.B.~thanks the physics department of Groningen university for its
hospitality. This work is supported by the European Commission TMR programme
ERBFMRX-CT96-0045, in which E.B.\ is associated to the university of
Utrecht. The work of K.B.\ is supported by the DFG.


\vspace{.5truecm}
\begin{thebibliography}{99}

\bib500 M.J.~Duff,
        {\em Supermembranes: the first 15 weeks}, {\em 
        Class.~Quantum Grav.}~{\bf 5}
        (1988) 189.

\bib600 E.~Bergshoeff, M.J.~Duff, C.N.~Pope and E.~Sezgin,
        {\em Supersymmetric supermem\-brane vacua and singletons},
        \PL199 (1987) 69.

\bib601 M.P. Blencowe and M.J. Duff, 
         {\em Supersingletons}, 
         \PL203 (1988) 229.

\bib602 H. Nicolai, E. Sezgin and Y. Tanii, 
        {\em Conformally invariant 
	supersymmetric field theories on $S^p \times S^1$ and super p-branes}, 
	\NP305 (1988) 483.

\bib603 M.~Flato and C.~Fronsdal,
         {\em J.~Math.~Phys.~}{\bf 22} (1981) 1100.

\bib620 P.~Claus, R.~Kallosh and A.~Van Proeyen,
        {\em M-5-brane and superconformal tensor multiplet in 6 dimensions},
        {\tt hep-th/9711161}.

\bib621 J.~Maldacena, {\em The large N limit of superconformal field
        theories and supergravity},
        {\tt hep-th/9711200}; {\em Wilson loops in large N field theories},
        {\tt hep-th/9803002}.

\bib622 R.~Kallosh, J.~Kumar and A.~Rajaraman,
        {\em Special conformal symmetry},
        {\tt hep-th/9712073}.

\bib812 S.\ Hyun, {\it U duality between three-dimensional and higher 
	dimensional black holes}, {\tt hep-th/9704005}.

\bib820 H.J.\ Boonstra, B.\ Peeters and K.\ Skenderis,
	{\em Duality and asymptotic geometries}, 
	\PL411 (1997) 59, {\tt hep-th/9706192};
	K.\ Sfetsos and K.\ Sken\-deris, {\em Microscopic 
	derivation of the Bekenstein-Hawking entropy formula for non--extremal 
	black holes}, {\tt hep-th/9711138}; H.J.\ Boonstra, B.\ Peeters 
	and K.\ Skenderis,
	{\em Branes and anti-de Sitter spacetimes}, {\tt hep-th/9801076}.

\bib623 S.~Ferrara and C.~Fronsdal,
        {\em Conformal Maxwell theory as a singleton field theory on $adS_5$, 
        IIB three branes and duality}, {\tt hep-th/9712239}; 
	{\em Gauge fields as composite boundary excitations},  
	{\tt hep-th/9802126};
        S.~Ferrara, C.~Fronsdal and A.~Zaffaroni, 
        {\em On N=8 Supergravity in $AdS_5$ and N=4 Superconformal
        Yang-Mills theory}, {\tt hep-th/9802203}; S.~Ferrara and A.~Zaffaroni,
        {\em N=1,2 4D Superconformal Field Theories and Supergravity in
        $AdS_5$}, {\tt hep-th/9803060}.
\newline
% \bib629 
	K.~Behrndt, {\em Branes in $N$=2, $D$=4 supergravity and the 
	conformal field theory limit}, {\tt hep-th/9801058};
\newline
%\bib624 
	P.~Claus, R.~Kallosh, J.~Kumar, P.~Townsend and A.~Van Proeyen,
       {\em Conformal Theory of M2, D3, M5 and D1+D5 Branes},
       {\tt hep-th/9801206};
\newline
%\bib625 
	N.~Itzhaki, J.M.~Maldacena, J.~Sonnenschein and S.~Yankielowicz,
        {\em Supergravity and the Large N Limit of Theories With
        Sixteen Supercharges},
        {\tt hep-th/9802042};
\newline
%\bib631 
	M.\ G\"unaydin and D.\ Minic, {\em Singletons, doubletons and
	M-theory}, {\tt hep-th/9802047};
\newline
%\bib626 
	G.T.\ Horowitz and H. Ooguri, {\em Spectrum of large $N$ gauge
	theory from supergravity}, {\tt hep-th/9802116};
\newline
%\bib627 
	E.\ Witten, {\em Anti de Sitter space and holography},
	{\tt hep-th/9802150};
\newline
	S.-J.\ Rey and J.\ Yee, {\em Macroscopic strings as haevy quarks
	in large $N$ gauge theory and anti-deSitter supergravity},
	{\tt hep-th/9803001};
\newline
        I.Ya.~Aref'eva and I.V.~Volovich,
        {\em On Large N Conformal Field Theories, Field Theories in
        Anti-de Sitter Space and Singletons},
        {\tt hep-th/9803028};
\newline
        S.~Minwalla, 
        {\em Particles on $AdS_{4/7}$ and Primary Operators on
        $M_{2/5}$ Brane Worldvolumes}, {\tt hep-th/9803053};
\newline
        O.~Aharony, Y.~Oz, Z.~Yin,
        {\em M Theory on $AS_p\times S^{11-p}$ and Superconformal
        Field Theories},
        {\tt hep-th/9803051};
\newline
        R.~Leigh and M.~Rozali,
        {\em The Large N Limit of the (2,0) Superconformal Field Theory},
        {\tt hep-th/9803068};
\newline
        M.~Bershadsky, Z.~Kakushadze and C.~Vafa,
        {\em String Expansion as large N Expansion of Gauge Theories},
        {\tt hep-th/9803076};
\newline
        E.~Halyo,
        {\em Supergravity on $AdS_{4/7}\times S^{7/4}$ and M Branes},
        {\tt hep-th/9803077}.

\bib626 S.\ Kachru and E.\ Silverstein, {\em 4d Conformal Field 
   	Theories and Strings on Orbifolds}, {\tt hep-th/9802183}.

\bib627 L.\ Castellani, A.\ Ceresole, R.\ D'Auri, S.\ Ferrara, P.\ Fr{\' e}
	and M.\ Trigiante, {\em $G/H$ M-branes and $AdS_{p+2}$ geometries},
	{\tt hep-th/9803039}		
	

\bib628 M.J.~Duff, H.~L\"u and C.N.~Pope, {\em $AdS_5 \times S^5$ Untwisted},
        {\tt hep-th/9803061}.

\bib845 G.W.~Gibbons and P.K.~Townsend, {\em Vacuum interpolation in
        supergravity via super p-branes}, \PRL71 (1993) 3754,
        {\tt hep-th/9307049}.

\bib830 I.\ Bakas, \PL343 (1995) 103, {\tt hep-th/9410104}.

\bib700 G.W.\ Gibbons, M.B.\ Green and M.J.\ Perry, {\it Instantons and
        7-branes in type IIB superstring theory}, \PL370 (1996) 37, 
        {\tt hep-th/9511080}, M.B.\ Green and M.\ Gutperle,
        {\it Effects of D--instantons},
	{\tt hep-th/9701093}.

\bib810 A.A.\ Tseytlin, 
	{\it Type IIB instanton as a wave in twelve dimensions},
	\PRL78 (1997) 1864, \PRL78 (1997) 1864, {\tt hep-th/9612164}.

\bib825 C.M.~Hull, \NP468 (1996) 113, {\tt hep-th/9512181}.
 
\bib826 C.~Vafa, \NP469 (1996) 403, {\tt hep-th/9602022}.

\bib701 J.H.~Schwarz, {\it Nucl.~Phys.~}{\bf B226} (1983) 269.

\bib800 P.~van Nieuwenhuizen and  A.\ Waldron, {\em On Euclidean spinors and 
	Wick rotations}, \PL389 (1996) 29 and \PL389 (1996) 29, 
	{\tt hep-th/9608174}.

\bib801 M.J.~Duff, H.~L\"u, C.N.~Pope and E.~Sezgin,
         {\em Supermembranes with fewer supersymmetries},
         \PL371 (1996) 206,
         {\tt hep-th/9511162}.

\bibitem{Strings97} A.A.~Tseytlin,
    {\em Interactions between Branes and Matrix Theories},
    to appear in the proceedings of the {\it Strings '97} meeting,
    {\tt hep-th/9709123}.

\bib850 E.\ Bergshoeff, M.\ de Roo, M.B.\ Green, G.\ Papadopoulos and
	P.K.\ Townsend, {\em Duality of type-II 7-branes and 8-branes}, 
         \NP470 (1996) 113,
          {\tt hep-th/9601150}.

\bib851 E.~Bergshoeff, B.~Janssen and K.~Behrndt, \PRD55 (1997) 3785;
        E.~Bergshoeff, {\em p-branes, D-branes and M-branes}, 
        {\tt hep-th/9611099}.

\bib847 R.~Kallosh and J.~Kumar, {\em Supersymmetry Enhancement of
        D-p-branes and M-branes},  \PRD56 (1997) 4934,
        {\tt hep-th/9704189}.

\bib855 A.A.~Tseytlin, \NP475 (1996) 149,
        {\tt hep-th/9604035}.

\bibitem{Tseytlin2} N.~Itzhaki, A.A.~Tseytlin and S.~Yankielowicz,
    {\em Supergravity Solutions for Branes localized within Branes},
    {\tt hep-th/9803103}.

\bib866 G.~'t Hooft, {\em Dimensional reduction in quantum gravity},
	{\tt hep-th/9310026}; L.\ Susskind, {\em The world as a hologram},
	{\it J.~Math.~Phys.}\ {\bf 36} (1995) 6377, {\tt hep-th/9409089}. 









\end{thebibliography}
\end{document}